# Improper Ferroelectricity in Helicoidal Antiferromagnet $Cu_3Nb_2O_8$

G. Sharma[1], J. Saha[1], S. D. Kaushik[2], V. Siruguri[2], and S. Patnaik[1a]

[1]*School of Physical Sciences, Jawaharlal Nehru University, New Delhi 110067, India*

[2]*UGC–DAE Consortium for Scientific Research, Mumbai Centre, R-5 Shed, BARC, Mumbai 400085, India*

$Cu_3Nb_2O_8$ is an unusual multiferroic compound that undergoes a series of magnetic ordering at low temperatures. Concurrent development of electric polarization has been reported at $T_{N1}$ ~25 K corresponding to a non-collinear helicoidal ordering (Phys. Rev. Lett. **107**, 137205 (2011)). But questions remain on the microscopic origin of such phenomena. In this communication, we report a detailed study of induced polar ordering in $Cu_3Nb_2O_8$ by performing polarization, magnetization, dielectric constant and heat capacity measurements over a broad range of temperature and magnetic field. The dielectric constant shows suppression in the magnetic field but it does not shift to lower temperature with the application of external field. The appearance of magneto-dielectric effect signifies the contribution of q-dependent magnetic correlation function with enhanced weightage in the presence of magnetic field. This magnetic correlation function is associated with the ferro-axial vector and our overall results support this mechanism for the observation of muliferroicity in $Cu_3Nb_2O_8$.

[a]*spatnaik@mail.jnu.ac.in*

Multiferroic materials exhibit interdependence of magnetization and electric polarization in the absence of external magnetic and electric field [1]. Such coexistence is rare and generally the electrical and magnetic ordering temperatures are far apart. The material in which both ordering parameters appear concurrently can have large magneto-electric coupling and thus technologically most promising [2]. Both collinear (e.g. $Ca_3CoMnO_6$, $Y_2CoMnO_6$) and non-collinear (e.g. $MnWO_4$, $TbMnO_3$, $Ni_3V_2O_8$) magnetic ordering can lead to such juxtaposition of mutually exclusive material properties [3-8]. But the microscopic reason could be quite different. While exchange-striction is responsible for such phenomena in collinear case, the emergence of electric polarization due to non-collinear magnetic order is generally ascribed to inverse Dzyaloshinskii-Moriya (DM) interaction or Katsura-Nagaosa-Balatsky (KNB) model [9]. The origin of exchange-striction (ES) lies in the in-equivalent inter-atomic forces in the unit cell where as that for spin-spiral structure is the spin orbit coupling (SOC). In the KNB model, electric polarization relates to the spin structure through the expression $\mathbf{P} \propto \mathbf{k}_{ij} \times (\mathbf{S}_i \times \mathbf{S}_j)$, where propagation vector $\mathbf{k}_{ij}$ connects two adjacent spin moments $\mathbf{S}_i$ and $\mathbf{S}_j$ [9]. Therefore, polarization $\mathbf{P}$ is confined to plane of spin rotation and is perpendicular to propagation vector of incommensurate spin orderings. In recent years, particularly after the discovery of multiferroicity in $MnI_2$ and $CuFeO_2$, it is believed that KNB model may not be appropriate for several helicoidally ordered spin-frustrated multiferroics [10,11].

Centro-symmetric $Cu_3Nb_2O_8$ is one such example where the cardinal principle of cycloidal multiferroics is violated and $\mathbf{P}$ is found to be almost perpendicular to the plane of rotation of spins [12]. Further, the driving force behind improper ferroelectricity in $Cu_3Nb_2O_8$ is controversial. While Johnson et al. have assigned this to chiral component of magnetic structure or the so called ferro-axial model [12], using detailed DFT calculations Li et al. have suggested that $\mathbf{P}$ is effectively derived from exchange-striction [13]. Moreover, since the reported saturation polarization is negligible ($17\times 10^{-4}$ μC/cm$^2$) and there have been questions about the genuineness of polar ordering. In this communication, we report synthesis and extensive characterization of polycrystalline $Cu_3Nb_2O_8$ and find the ferro-axial model [12] to be more appropriate towards explaining our data. We confirm that ferroelectricity in $Cu_3Nb_2O_8$ emerges simultaneously below an incommensurate magnetic transition around $T_{N1}$= 25K [12]. The corresponding magnetic structure is attributed to a generic helicoidal spin arrangement. The appearance of cross coupling nature of magnetic and electric ordering is studied through detailed electrical polarization, dielectric constant and heat capacity measurements as a function of temperature and magnetic field.

**Experiments**

The polycrystalline samples of $Cu_3Nb_2O_8$ were prepared by solid state reaction method. The starting materials CuO and $Nb_2O_5$ were taken in stoichiometric ratio. It was ground and pelletized by applying hydrostatic pressure of 5 tonnes. First sintering was done at 800°C for 36 hour and next was done at 950°C for 40 hour. Powder X-ray diffraction (XRD) pattern was performed using PANalytical X'Pert PRO at room temperature and Rietveld refinement was done with GSAS software [14]. Neutron diffraction (ND) was carried out on powder sample using multiple sensitive detectors (PSD) Focussing Crystal Diffractometer (FCD) set up by UGC-DAE Consortium for Scientific Research Mumbai centre at the National Facility for Neutron Beam Research (NFNBR), Dhruva reactor, Mumbai (India) at a wavelength of 1.48Å. The sample was placed in vanadium can which was then exposed to neutron beam for neutron diffraction pattern. The data were analysed using Rietveld method by using the FULLPROF program [15]. The sample pellet was painted with silver paste on both sides for the capacitance measurement under various temperature and magnetic field conditions.

Similar configuration was employed for the pyroelectric current measurement. A QUADTech 1920 Precision LCR meter was used for the temperature dependence of magneto-capacitance measurements. To obtain temperature and magnetic field dependence of polarization, pyroelectric current was measured using 6514 Keithley electrometer while heating the sample at a rate of 3 K per minute. Electric polarization was obtained by integrating pyroelectric current over time. Prior to the measurements, the samples were poled with an applied external electric field ~200kV/m. The dielectric and pyroelectric current were measured in cryogen free magnet (CFM) system. The heat capacity and magnetic property measurement was done in *Cryogenic* PPMS and *Quantum Design* MPMS.

**Results and Discussions**

Figure 1(a) and 1(b) depict room temperature XRD and ND patterns. The data confirm that single phase $Cu_3Nb_2O_8$ has been synthesized and both patterns could be satisfactorily indexed in centro-symmetric triclinic structure (space group $P\bar{1}$) [12, 16]. The cell parameters were calculated by using both XRD and ND data and the values from ND data are as follows; **a**=5.1859 Å, **b**=5.4871 Å, **c**=6.0178 Å, α=72.542°, β=83.498° and γ=65.701°. These values are consistent with earlier reports [12, 16]. The cell parameters determined from XRD were slightly smaller (e.g. **a** = 5.1795 Å), but position of light ions such as oxygen could be ascertained with greater confidence from neutron data. .

The unit cell of $Cu_3Nb_2O_8$ is shown in Fig. 1(c). $Cu^{2+}$ has two in-equivalent atomic sites. $Cu^{2+}(1)$ is at the interior of the unit cell within a bi-pyramidal configuration where as $Cu^{2+}(2)$ is at the corners of the unit cell each coordinated to four oxygen atoms. The atomic position coordinates, $Cu^{2+}$-O bond lengths and $Cu^{2+}(1)$-O-$Cu^{2+}(2)$ bond angles derived from the room temperature powder ND pattern are given in Table I. The bond length of $Cu^{2+}(1)$-O ranges from 1.898 to 2.708 Å and that for $Cu^{+2}(2)$-O ranges from 1.888 to 2.783 Å. Niobium ($Nb^{5+}$) ions on the other hand, are adjacent to $Cu^{2+}(1)$ ions and are octahedrally surrounded by six oxygen atoms [12, 16].

In Fig 2a and b we plot DC magnetization (M), and heat capacity ($C_P$) as a function of temperature. Two consecutive transitions appear at 25 K and 27 K in the heat capacity data. Magnetization versus temperature scan confirms onset of anti-ferromagnetic transition at ~ 27 K. As seen in the inset Fig 2a, a divergence in the derivative of magnetic susceptibility around $T_{N1}$ is observed at 25 K. Further, the broad peaks in $C_P/T$ corresponding to the magnetic transitions get suppressed and shifted towards lower temperature at H = 5 and 10 Tesla. The baseline of specific heat (below 22 K) of 0.97 J/mole-$K^2$ is associated with development of short range magnetic correlation that relate to collective contribution of lattice and spin [14]. In the inset of Figure 2b, heat capacity $C_P$ is plotted over a broader temperature range and we can observe deviation around 60 K as well, the reason for which is yet to be ascertained. In Figure 3, we study the magnetization plots as a function of temperature and external magnetic field in detail. The inverse magnetic susceptibility follows Curie-Weiss law in the paramagnetic state (Susceptibility $\chi=C/(T+\theta)$, C and $\theta$ being the Curie constant and Weiss temperature respectively) down to 60 K from room temperature (inset 3**a**). Curie-Wiess fitting yields C= 2.9 emu-K/mole$^{-1}$ Oe$^{-1}$ and the effective paramagnetic moment $\mu_{eff}$=1.6$\mu_B$ per $Cu^{2+}$ ion. The magnitude of $\theta$ is 157 K and the intercept in the negative axis reflects anti-ferromagnetic correlations. The degree of frustration, defined as the ratio between $\theta$ and $T_N$, is estimated to be ~6 which is similar to values reported for elliptical spin spiral $MnWO_4$ [17]. Such high ratio shows relevance of second nearest neighbour interaction in $Cu_3Nb_2O_8$. As seen in inset of Fig 3b, the magnetization versus magnetic field scan at 5K shows hysteretic behaviour, indicative of canted spin arrangement. Note that no strong deviation in magnetization is observed corresponding to $T_{N1}$. Further, the deviation at 60 K observed in the heat capacity measurement is also corroborated in the magnetization plot.

Next we discuss the temperature dependence of electric polarization in zero and 7 T external magnetic field (Fig 4). It is evident that the polarization starts to develop at $T_{N1}$ (25 K) which saturates to ~7μC/m$^2$ at low temperature. It is reported that maximum saturation polarization of 17μC/m$^2$ [12] is achieved along (0, 1, 0) plane in the single crystalline sample, so in our polycrystalline sample 7μC/m$^2$ represents average over various grain orientation. Inset of Fig 4a shows temperature dependence of measured pyroelectric current that exhibits a sharp anomaly near $T_{N1}$. Since the magnitude is small, it could very well be due to experimental artefacts such as contribution of thermally induced trap charges [18]. To confirm genuine ferroelectricity, the reversibility of polarization direction was checked by changing direction of applied electric field during poling. Figure 4b confirms directional change of polarization with poling field sign. The method of measurement of pyroelectric current under small thermal cycling was also carried out [18]. Inset 4c shows the plot where both sample temperature and pyroelectric current are measured as a function of time under thermal cycling (below $T_{N1}$). We note that the pyroelectric current direction follows concurrently with the heating/cooling cycle. This result confirms that the improper ferroelectricity in $Cu_3Nb_2O_8$ is due to genuine polar activity [18].

The magnetic field dependence of dielectric constant as a function of temperature is shown in Fig 5(a). With the application of magnetic field, the dielectric constant decreases. The magneto-dielectric coupling parameter [(ε(0T) - ε (5T)) / ε (0T)] in the vicinity of transition temperature is plotted in Fig 5(b). Corresponding to temperature dependent specific heat and pyroelectric current data, a sharp peak in dielectric constant is also observed at $T_{N1}$ (Fig 5(a)). The variation in dielectric constant remains insensitive to other magnetic phases (namely at $T_{N2}$~27 and $T_{N3}$~60K). To bring in a comparison, magneto-dielectric parameters in collinearly ordered multiferroics, are reported to be around 28% in $GdFeO_3$ [19] and around 5% in $Ca_3Co_2O_6$ [20]. On the other hand, for non-collinearly ordered anti-ferromagnets, it acquires a value 0.1% ($MnWO_4$ [17]), 0.3% ($Ni_3V_2O_8$ [21])). The small coupling parameter is generally suggestive of non-collinear mechanism but firm inferences cannot be drawn in a polycrystalline pellet. Further, the appearance of magneto-dielectric effect is due to variation in magnetic correlation function $I(T)$ [22, 23] and the effective dielectric constant $\epsilon$ can be expressed as

$$\epsilon = \frac{\epsilon_o}{1+2\epsilon_o I(T)} \quad [1]$$

Here, q-dependent spin-spin correlation function is $I(T) = \sum_q g(q) <M_q M_{-q}>(T)$ and $\varepsilon_0$ is the dielectric constant measured in the absence of magnetic field. Vectors q and -q are the spin states at adjacent sites (i, j) i.e., $S_i$ and $S_j$. The parameter g(q) and $<M_qM_{-q}>$ are q-dependent coupling constant and thermal average of instantaneous spin-spin correlation respectively [24]. As observed in Fig 5, in $Cu_3Nb_2O_8$ the effective dielectric constant diverges at ~ 25K but it is suppressed with the application of external magnetic field. We note that there is a difference between $Ni_3V_2O_8$ (or $MnWO_4$) and $Cu_3Nb_2O_8$ in the magneto-dielectric data. The anomaly in dielectric constant shifts to lower temperature under application of magnetic field for $Ni_3V_2O_8$ which is not the case with $Cu_3Nb_2O_8$ [5, 25, 26].

While a microscopic correlation of magnetic structure with the magneto-electric coupling would require a detailed magnetic field dependent neutron diffraction experiment, in the following we present a qualitative understanding of our data. Existing analysis suggests that $Cu_3Nb_2O_8$ belongs to a rare class of paramagnetic group where macroscopic structural rotation is allowed and this gets coupled unto magnetically ordered chiral phase below $T_{N1}$. This has been described as the ferro-axial model [12]. In such a case, the nonzero chiral term ($\sigma_{ij}$) induces the macroscopic polar ordering **P** = γσ**A** where γ and **A** are coupling constant and macroscopic axial vector respectively [12, 29]. The appearance of anomaly in dielectric measurement near magnetic transition reflects on the strength of coupling between lattice and spin and the suppression of dielectric anomaly in the presence of magnetic field confirms a local correlation between the electric dipole and magnetic ordering [32]. A general feature of SOC driven multiferroics, as evidenced in TbMnO$_3$, Ni$_3$V$_2$O$_8$, and MnWO$_4$ is a shift in dielectric anomaly temperature at relatively low fields. This is not seen in Cu$_3$Nb$_2$O$_8$. A third possibility could be relevant, as discussed regarding giant ferroelctricity observed in CaMn$_7$O$_{12}$ [29]. In this case, the magnitude of polarization is assigned to exchange-striction but the direction of polarization is described by chilarlity of helical magnetic order. But here the exchange-striction is possible because mixed valent states of Mn$^{3+}$ and Mn$^{4+}$. There is no such possibility in Cu$_3$Nb$_2$O$_8$. In essence, our magneto-dielectric data supports the ferro-axial coupling based mechanism.

In summary, we report on the synthesis and detailed characterization of multiferroic properties of spin frustrated anti-ferromagnet Cu$_3$Nb$_2$O$_8$. The spontaneous ordering of magnetic and induced ferroelectric parameters occurs simultaneously in Cu$_3$Nb$_2$O$_8$ around one of its anti-ferromagnetic transition temperature T$_{N1}$ ~ 25K. Although the magnitude is small, genuine improper ferroelectricity is established. Anomaly in dielectric constant around T$_{N1}$ and its suppression in the presence of external field indicates strong spin/charge coupling to magnetic ordering. The spin chiral-ferroaxial vector correlation is modified by magnetic field and relates to the magneto-dielectric data. This magneto-dielectric coupling strongly favours ferro-axial mechanism for the observed improper ferroelectricity in Cu$_3$Nb$_2$O$_8$.

**Acknowledgement**

Authors would like to acknowledge for very helpful conversation with Prof. A. K. Rastogi (JNU) and Dr. Y. Bugoslavsky (*Cryogenic* ltd.). DST, Government of India, is acknowledged for funding the SQUID at IIT Delhi. We thank AIRF, JNU for access to PPMS. GS and JS acknowledge University Grant Commission (UGC) and Council of Scientific and Industrial Research (CSIR), India for financial support.

Table I. Structural parameters (atomic position, bond length, and bond angle) obtained from the Rietveld refinement of ND pattern of $Cu_3Nb_2O_8$.

| Atom | x | y | z |
|------|---|---|---|
| Cu1 | 0.000 | 0.000 | 0.000 |
| Cu2 | 0.467(8) | 0.072(3) | 0.349(8) |
| Nb | 0.221(4) | 0.539(1) | 0.650(5) |
| O1 | 0.230(1) | 0.205 | 0.897(3) |
| O2 | 0.262(1) | 0.756(6) | 0.831(3) |
| O3 | 0.365(5) | 0.759(8) | 0.365(8) |
| O4 | 0.174(3) | 0.307(1) | 0.399(5) |

| Bond | Length (Å) | Bond | Angle |
|------|------------|------|-------|
| Cu(1)-O(1) | 1.898×2 | Cu(1)-O1-Cu(2) | 42.381 |
| Cu(1)-O(2) | 1.933 ×2 |  | 137.619 |
| Cu(1)-O(3) | 2.7086×2 |  | 91.951 |
|  |  |  | 40.005 |
| Cu(2)-O(1) | 1.9907 | Cu(1)-O2-Cu(2) | 42.177 |
|  | 2.2351 |  | 137.823 |
|  |  |  | 40.977 |
| Cu(2)-O(2) | 1.9796 | Cu(1)-O3-Cu(2) | 78.313 |
| Cu(2)-O(3) | 1.8888 |  | 101.687 |
|  |  |  | 119.494 |
| Cu(2)-O(4) | 2.7837 | Cu(1)-O4-Cu(2) | 141.971 |
|  | 1.9174 |  | 80.227 |

**Figure Captions:**

**Fig 1.** (Color online) (a) Room temperature X-ray diffraction pattern of $Cu_3Nb_2O_8$ with its Rietveld refinement. (b) Room temperature powder neutron diffraction pattern. The observed data (black circle), calculated (red line), difference (blue line) and Bragg position (tick mark) are shown (c) Schematic triclinic crystal structure of $Cu_3Nb_2O_8$. $Cu^{2+}$, $Nb^{5+}$ and $O^{2-}$ are shown in green, blue and red respectively.

**Fig 2.** (Color online) (a) Magnetization vs temperature at 0.1T magnetic field. Inset (a) shows temperature derivative of magnetization which indicates that ferroelectric ordering at $T_{N1}$ has magnetic origin (b) $C_P/T$ measurement carried out at constant filed of H = 0T, 5T and 10T. Two anomalies appear in the specific heat measurement. The peaks at 27K and 25K is corresponds to two successive anti-ferromagnetic transitions. Heat capacity as a function of temperature is shown in inset of (b) in a broader temperature scale.

**Fig 3.** (Color online) Temperature dependence magnetic behaviour of $Cu_3Nb_2O_8$ under 0.1T magnetic field, Inset (a) shows inverse of magnetic susceptibility. Inset (b) shows hysteresis in magnetization at T = 5K.

**Fig 4**. (Color online) Temperature dependence of electric polarization **P** (Poling field E = 200kV/cm). Inset (a) shows measured pyroelectric current as a function of temperature. This data are used to derive electrical polarization. Inset (b) elucidates reversibility of polarization with poling electric field. Inset (c) shows response of pyroelectric current under thermal cycling.

**Fig 5**. (Color online) (a) Temperature dependence of dielectric constant in presence of applied external magnetic field of strength H = 0, 5, 7 Tesla at frequency 5023 Hz. (b) Temperature dependence of magneto-dielectric parameter across the ferroelectric transition.

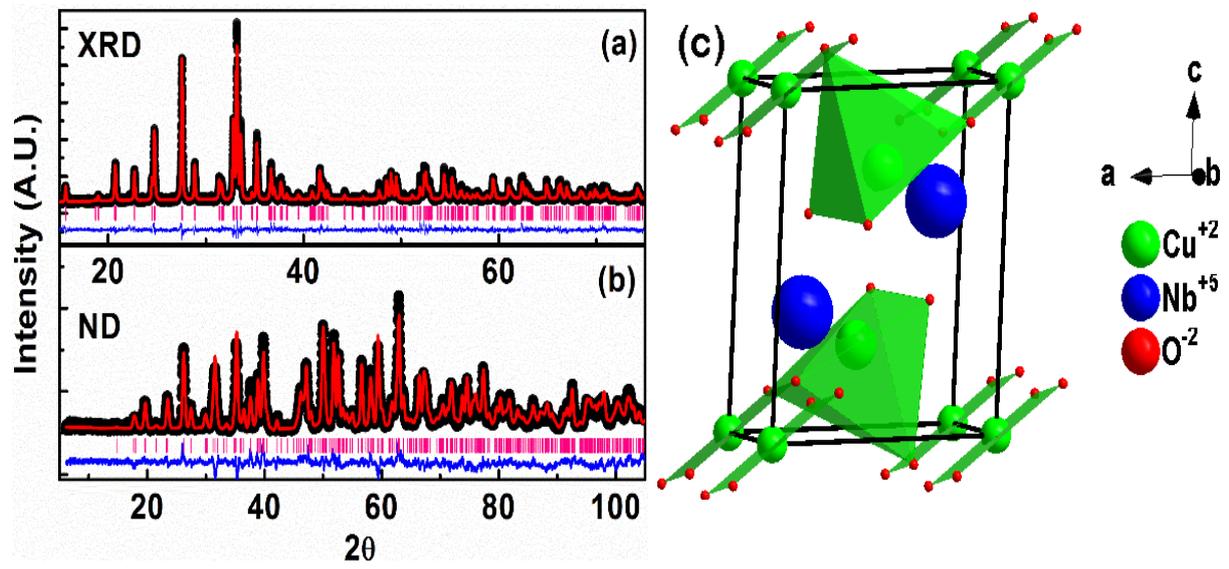

**FIG. 1**

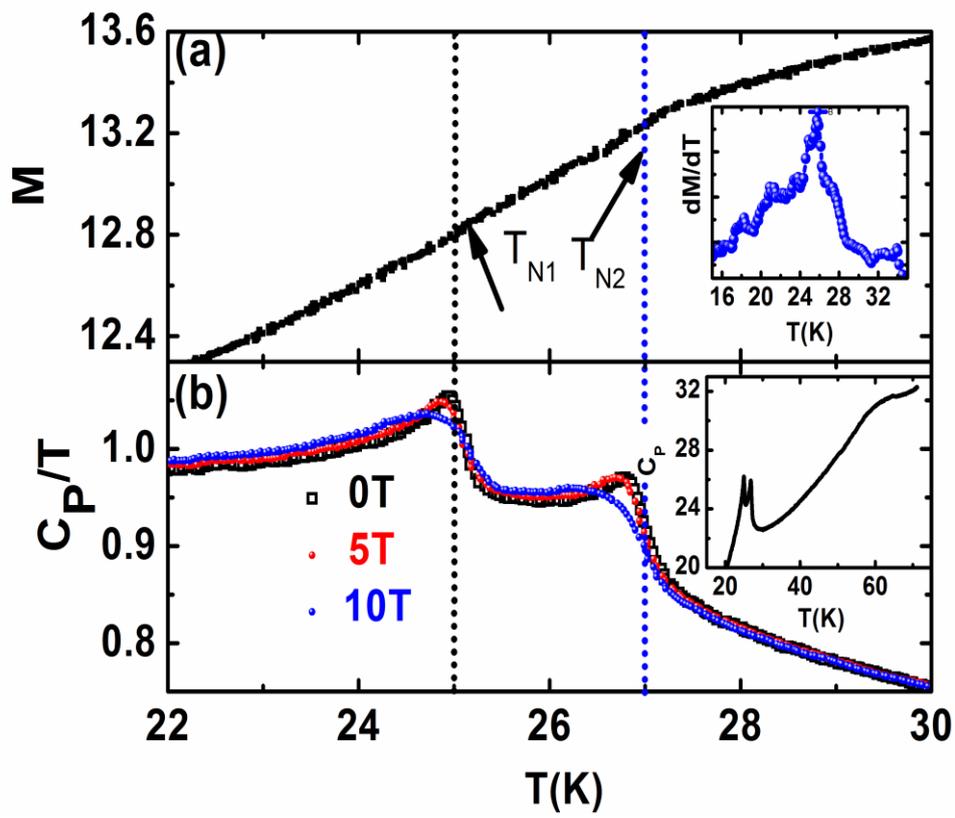

**FIG. 2**

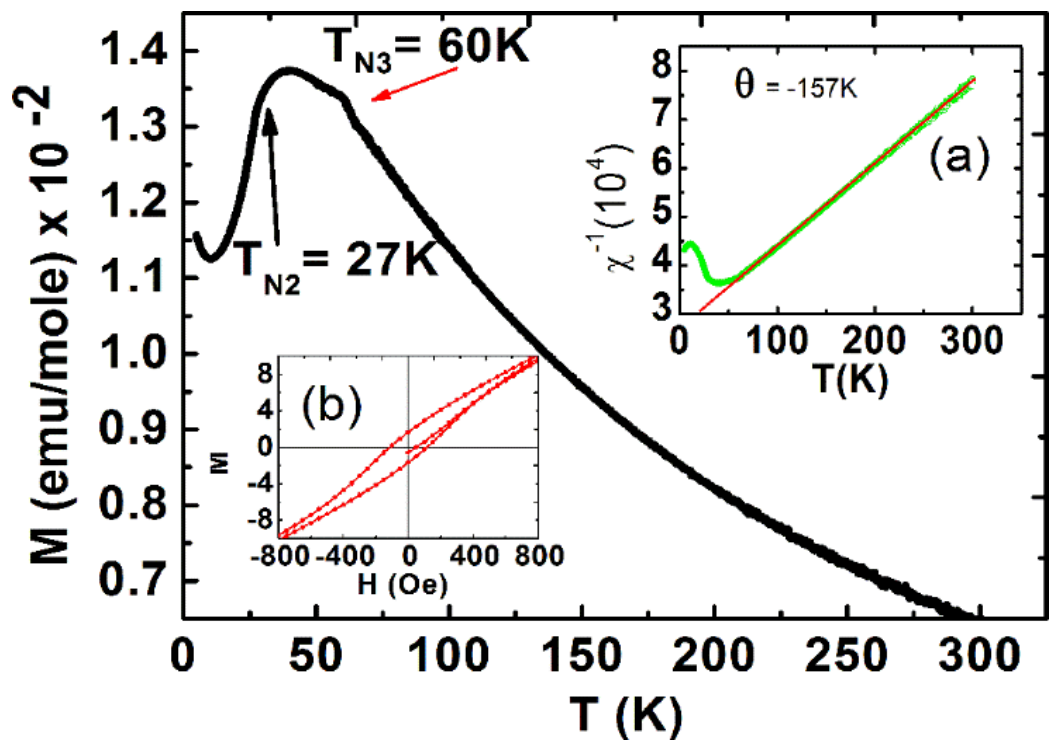

**FIG. 3**

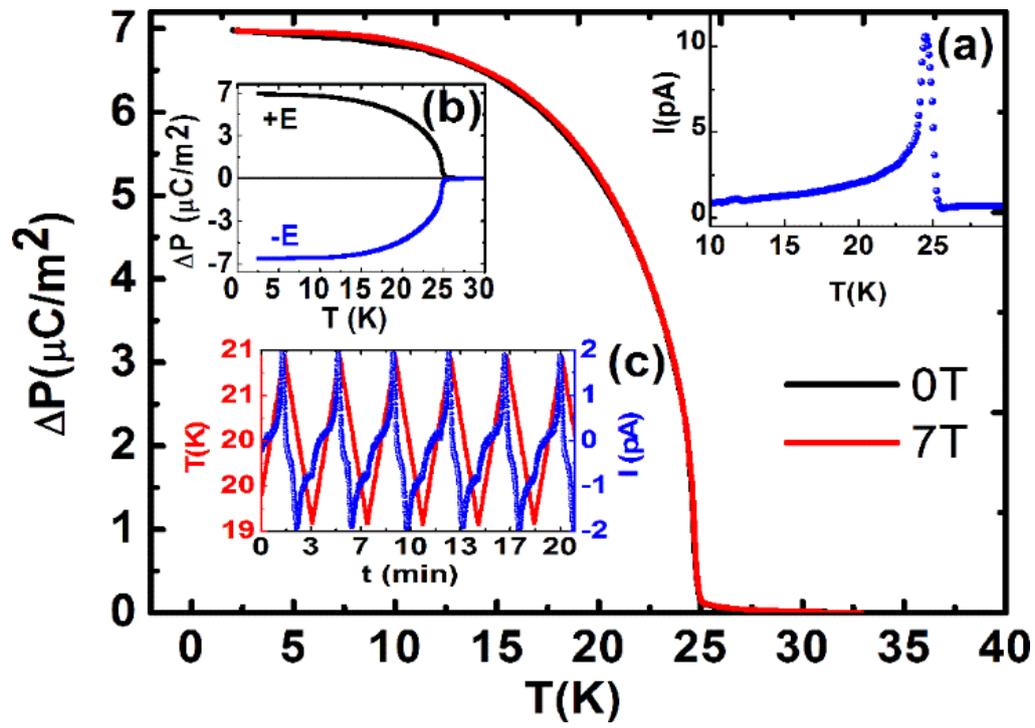

FIG. 4

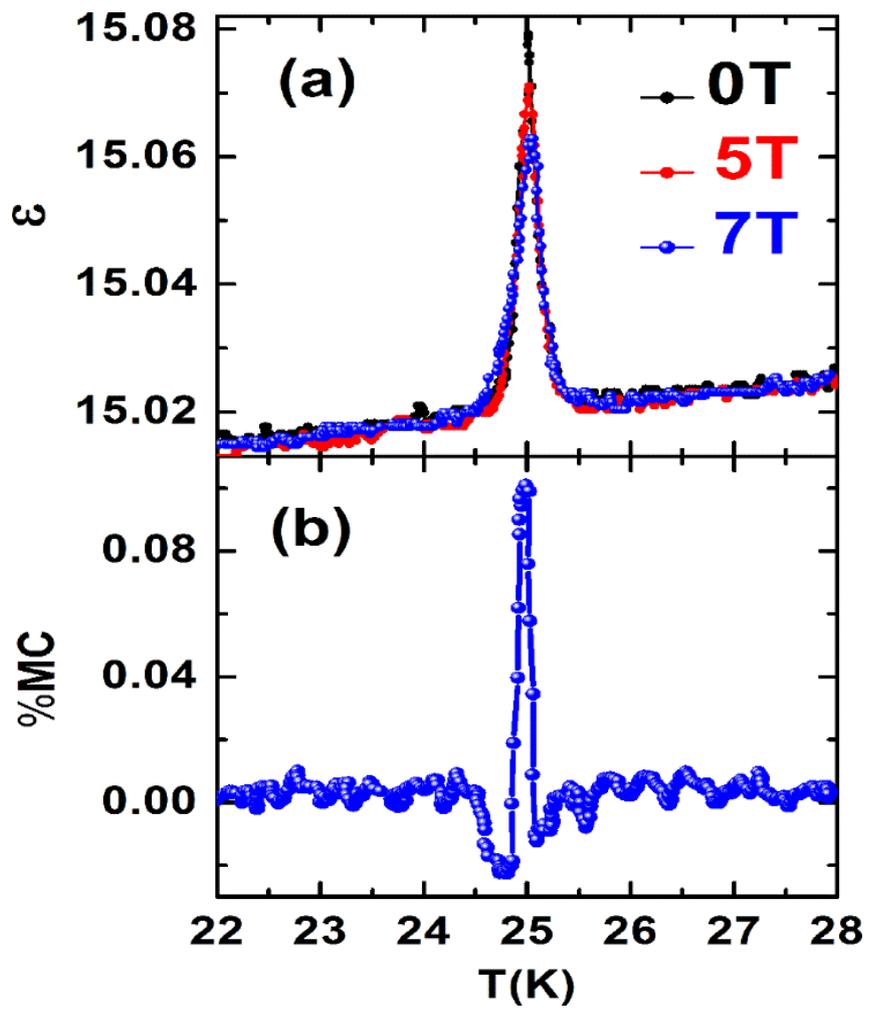

**FIG. 5**